\documentclass[11pt]{amsart}
\pdfoutput=1

\usepackage[utf8]{inputenc}

\usepackage[margin=3.25cm]{geometry}

\usepackage{amssymb, amsmath, amsthm, amsfonts}
\usepackage{graphicx}
\usepackage{color}
\usepackage{mathrsfs}
\usepackage{physics}
\usepackage{mathtools}
\usepackage{tikz}
\usepackage{faktor}
\usepackage{tikz-cd}
\usepackage{bm}
\usepackage[hidelinks]{hyperref}
\usepackage{csquotes}

\DeclareRobustCommand{\SkipTocEntry}[5]{}

\theoremstyle{remark}

\theoremstyle{remark}

\usepackage[UKenglish]{isodate}
\usepackage[UKenglish]{babel}

\DeclareMathOperator{\E}{\mathbb{E}}

\title[Regarding Flows Under the Free Energy Principle]{Regarding Flows Under the Free Energy Principle: \\ A Comment on ``How Particular is the Physics of the Free Energy Principle?" by Aguilera, Millidge, Tschantz, and Buckley}

\author{Dalton A R Sakthivadivel}
\address{\parbox{\linewidth-12pt}{Departments of Mathematics, Physics and Astronomy, and Biomedical Engineering, Stony Brook University, Stony Brook, NY, USA, 11794-3651}}
\address{VERSES Research Lab and Spatial Web Foundation, Los Angeles, CA, USA, 90016}
\email{dalton.sakthivadivel@stonybrook.edu}
\urladdr{https://darsakthi.github.io}

\date{\today}

\let\oldtocsection=\tocsection

\let\oldtocsubsection=\tocsubsection

\renewcommand{\tocsection}[2]{\hspace{0em}\oldtocsection{#1}{#2}}
\renewcommand{\tocsubsection}[2]{\hspace{19.5pt}\oldtocsubsection{#1}{#2}}

\begin{document}

\maketitle

\begin{abstract}

In a recent technical critique of the free energy principle (FEP) due to Aguilera-Millidge-Tschantz-Buckley, it is argued that there are a number of instances where the FEP\textemdash as conventionally written, in terms of densities over states\textemdash is uninformative about the dynamics of many physical systems, and by extension, many `things.' In this informal comment on their critique, I highlight two points of interest where their derivations are largely correct, but where their arguments are not fatal to the FEP. I go on to conjecture that a path-based formulation of the FEP has key features which restore its explanatory power in broad physical regimes. Correspondingly, this piece takes the position that the application of a state-based formulation of the FEP is inappropriate for certain simple systems, but, that the FEP can be expected to hold regardless.

\end{abstract}

\date{\today}



\section{Two problems raised by the target paper}

We briefly recount two primary conclusions of the target paper. One is presented by \cite{aguilera2021} as a counterexample to the generality of certain solenoidal descriptions of flow, and the other, to the claim that free energy gradients are informative about linear systems.

\subsection{No solenoidal coupling is no strict sign of system-ness}\label{solenoid-problem} One of the primary results of \cite{aguilera2021} is that the lack of solenoidal couplings between external and other states, which one (I emphasise preemptively: not completely faithful) reading of the FEP might suggest is a general feature of physical systems, is rare. The rationale for this is introduced in Assumption 1 of the paper and relevant results are given in \cite[section 3.1]{aguilera2021}, in Figure 4 and equations 30 through 37. The presence of solenoidal couplings in linear systems (i.e., the absence of sparse coupling), along with the extrapolation that they will be present in more complicated, non-linear systems, is suggested as an indication that the FEP does not hold for many systems which it might seek to describe. As it is argued, the consequence of this is that sparse coupling is a sufficient, but not necessary, sign of system-ness. Under this general reading of the FEP, their result is initially in conflict with the way systems and system-ness are described. 

\subsection{Free energy is uninformative about the flow of systems under linear fluctuations}\label{linear-problem}

In \cite[section 3.2]{aguilera2021}, the authors discuss how the marginal flows of internal and external states on free energy and surprisal gradients, respectively, are uniformative about a linear system's behaviour. In particular, Figure 5 of the target paper shows a poor correlation between system dynamics and marginal dynamics; this might lead one to believe the FEP is a poor descriptor of the system. As such, the very application of the FEP to linear systems is called into question by the authors.

\section{Two possible path-based solutions}

Now, I propose some potential solutions to the issues raised by \cite{aguilera2021}, mainly in the language of probabilities over the flows\textemdash the paths of evolution\textemdash of a noisy system. Since the FEP is naturally path-based \cite{afepfapp, simpler, barp}, it is fitting that the problems posed can be circumvented by returning to paths. As such, I posit that the above arguments are not fatal to the FEP.

One might expect path-based descriptions to be superior to state-based ones for one simple reason: maximum calibre is asymptotically maximum entropy. This laconic statement hides a more profound observation that systems in non-equilibrium regimes, where fluctuations do not obey the ensemble properties typical of a system, are described well by probabilities over the possible trajectories of the system instead \cite{udo, touchette}. This is referred to as the principle of maximum calibre by Jaynes \cite{jaynes1985macroscopic, dill}. Non-stationary diffusion processes, non-equilibrium systems whose sampling statistics change from time-point to time-point, are also well-described by maximum calibre \cite{ken2}. 
Besides maximum calibre, the evolution of diffusing systems in non-stationary regimes is famously solved backwards by the path integral \cite{oksendal, moral, glimm}. In \cite{afepfapp}, it is stated throughout that generalised coordinates of motion are the proper, dynamical setting for the FEP, and that connections between the FEP and the fluctuation theorems of stochastic thermodynamics are natural. In the recent papers \cite{simpler} and \cite{barp}, this is used to great effect. Most importantly, the problems summarised in Sections \ref{solenoid-problem} and \ref{linear-problem} here are indications of the fact that a state-based formulation is simply not an apt description of certain systems. 

Here, I sketch out two ways in which the previous issues\textemdash and implicitly, many more questions surrounding the FEP\textemdash may be resolved, by looking at paths as opposed to states. I am grateful to Karl J Friston for helpful discussions about these ideas, and for sharing preliminary work writing the FEP in terms of dynamical trajectories explicitly.

\subsection{No solenoidal coupling is a special case of path-wise independence}

The key idea behind the FEP, of which sparse coupling is but a shadow, is that of \emph{individuation}. In \cite{aguilera2021} it is demonstrated that, amongst linear systems, the sorts of conditional independencies underlying the Markov blanket are difficult to construct. It is inferred that they will be rare in general, and that sparse coupling only holds for particular systems. I contend that the Markov blanket as it is typically presented is a special case of individuation along the dynamical evolution of a system. As soon as one commits to an idea of a `thing' in the context of the FEP, one commits to an individuation of that thing from its environment, such that the system evolves along its own set-points and distinguishes its structure from that of the external milieu surrounding it. In this case, the flow of a system is conditionally independent of its environment. 

A concrete way in which this construction should avoid problems raised by \cite{aguilera2021} is that the stereotypical structure of the Hessian should become unnecessary, and these conditional independencies should be made obvious from either the Jacobian of the system, or the probabilities over system-like trajectories given environmental dynamics. With no density over states there is no wondering about solenoidal flows and conditional independencies; these go away when states go away. A lack of solenoidal coupling may describe specific systems in an interesting way, but ought to be a special case of this more general construction. 

\subsection{Linear systems match, but do not track, minima of free energy}

In the target paper, the example of a linear system is used, where it is shown that these systems do not exhibit interesting or informative behaviour under gradient flows of the free energy. The authors infer that free energy is not a useful quantity to describe simple systems, and that it will be even more difficult to use the FEP insightfully in the more complicated systems which are of interest to the FEP. 

Here, the paths-based view clarifies the role of non-stationary dynamics in the FEP. Under linear fluctuations, the system dissipates to a fixed point $\mu^*$ on average. The most likely state that the FEP references is trivial, and there is simply nothing interesting \emph{about} the free energy of this system\textemdash the system engages in mode-matching by dissipating to $\mu^*,$ but does not engage in any dynamical mode-tracking, and hence the flow on free energy is trivial. To that end, note that Figure 5 of \cite{aguilera2021} demonstrates precisely this: the evolution of a small number of samples is not representative of the FEP. 

Writing flows along free energy gradients in the language of information geometry (where beliefs typically change), it becomes apparent that the FEP does apply to systems with stationary sufficient statistics\textemdash but by and large, has nothing interesting to say about them. 
Instead, these sorts of marginal flows appear to have a home in maximum calibre, whose marginal densities along the flow of a mode correspond to specific beliefs. Suppose the trajectory taken by the system satisfies equation 3.6 in \cite{parr} or equation 22 in \cite{aguilera2021}, i.e.,  
\begin{equation}\label{constraint}
\langle \mu(t) \rangle_{\omega} = \int_0^t f_\mu(\bm\mu_s, b) \dd{s},
\end{equation}
with $\langle \cdots \rangle_\omega$ being an expectation with respect to realisations of noise $\omega$. For a sample path $\gamma$, one can denote \eqref{constraint} identically as
\[
\E_{p(\gamma)} [\gamma] = C(t).
\]
If $\pi$ is the space of paths and $\gamma = \{\mu_0, \ldots, \mu_n\}$ is a path therein, the idea is to construct mode-tracking by constraining maximum calibre in the following way:
\[
- \int_{\pi} \ln\{ p(\gamma) \} p(\gamma) \dd{\gamma} - \lambda(t) \left( \int_{\pi} \gamma \, p(\gamma) \dd{\gamma} - C(t) \right),
\]
where $\gamma$ is a trajectory of states, and the constraint is \eqref{constraint}. In other words, that the ensemble average (i.e., the average with respect to a density over realisations of sample paths) of $\gamma$ equals a particular function $C(t)$. Assuming independence across steps in time simply for notational ease, the resultant density, 
\begin{equation}\label{calibre-density}
\exp{-\lambda_{0, \ldots, n}(\mu_0, \ldots, \mu_n) },
\end{equation}
is the concatenation of several individual beliefs (each of which is parameterised by some mode) according to the laws of joint probability:
\[
\exp{-\lambda(n) \mu(n) }\exp{-\lambda(n-1) \mu(n-1) }\ldots \exp{-\lambda(0)\mu(0) }.
\]
Mapping this flow into a statistical manifold, the flow appears to describe an inference process where the belief at every $\mu_t$ changes. In particular, each such density has $\bm\mu_t$ as a parameter, given as its sufficient statistic. Conjecturally, a process modelled by \eqref{calibre-density} can be read as a system which performs dynamical inference about where the mode may be at a future time-step, and then flows towards that mode. 

Since this appears to be equivalent to the FEP formulation in \cite{afepfapp} and \cite{parr}, it is clear that the FEP is not indeed a very useful treatment for linear systems, where the ensemble average of $\gamma$ is the constant function equal to $\mu^*.$  That is, it is clear that the interesting, mode-tracking behaviour of non-linear systems which is key to the utility of the FEP is absent from this description. To that end, it is likely that, in virtue of aiming towards `things' (defined as individuated systems with a particular partition) and not arbitrary systems, the FEP is naturally most informative for non-linear or chaotic systems, where it has been used to effect \cite{stoch-chaos}.

Along the same lines, only the \textit{fluctuations} of linear systems contain any non-trivial dynamics, which are washed away by the mode-matching under the FEP. That the density-over-states formulation assumes such a limit\textemdash an idealised non-equilibrium steady state, valid for an ensemble of systems and not a particular realisation of a system\textemdash rather than using probabilities of trajectories, manifests this uninformativeness. Probabilities over trajectories, meanwhile, could include transients or few-body systems \cite{dill}, and are also better suited to dynamical modes.

As such, the FEP has little of interest to say about linear systems, because linear systems are not interesting. This sort of reformulation should resolve the tension between the indicated papers and \cite{aguilera2021}, by demonstrating more clearly that the FEP is not, after all, useful for linear systems; but that this is by design.

\section{Closing remarks}


It must be reasserted that this comment takes a middle way between dismissing and sermonising the FEP: the foregoing points of critique, first raised in \cite{aguilera2021}, are mathematically correct. However, I do not believe they invalidate the FEP\textemdash nor, one presumes, do the authors of \cite{aguilera2021}, who instead produce a working model of the FEP and point out what is not possible within that model. Rather than show that the FEP \emph{per se} is incorrectly formulated, I view \cite{aguilera2021} as demonstrating that the FEP needs further unravelling\textemdash for instance, returning to the paths-based constructions intimated throughout \cite{afepfapp}\textemdash to make the nature of the tools it offers clear. 

\bibliographystyle{alpha}
\bibliography{main}

\end{document}